\documentclass[aps,10pt,superscriptaddress,twocolumn]{revtex4-1}
\usepackage{graphicx}
\usepackage{epstopdf}
\usepackage{amssymb}
\usepackage{mathrsfs}
\usepackage{amsmath}
\usepackage{xcolor}
\usepackage{latexsym}
\usepackage{amsfonts}
\usepackage{hyperref}
\usepackage{subfigure}
\usepackage{wasysym}
\usepackage{dcolumn}
\usepackage{bm}
\usepackage[percent]{overpic}
\usepackage{natbib}
\usepackage{booktabs}
\usepackage{gensymb}
\usepackage{lineno}

\begin{document}

\selectfont
\title{A colloid approach to self-assembling antibodies}

\author{Nicholas Skar-Gislinge}\email[]{present address: Copenhagen Business School, Porcelaenshaven 18B, 2000 Frederiksberg, Denmark }
\affiliation{Physical Chemistry, Department of Chemistry, Lund University, SE-221 00 Lund, Sweden}
\author{Michela Ronti} \email[]{present address: Computational Physics, University of Vienna, Sensengasse 8, 1090 Vienna, Austria}
\affiliation{Department of Physics, Sapienza Universit\`a di Roma, Piazzale Aldo Moro 1, 00185 Rome, Italy}
\author{Tommy Garting}
\affiliation{Physical Chemistry, Department of Chemistry, Lund University, SE-221 00 Lund, Sweden}
\author{Christian Rischel}
\affiliation{Novo Nordisk A/S, DK-2760 Malov, Denmark}
\author{Peter Schurtenberger}
\affiliation{Physical Chemistry, Department of Chemistry, Lund University, SE-221 00 Lund, Sweden}
\author{Emanuela Zaccarelli}\email[]{emanuela.zaccarelli@cnr.it}
\affiliation{Institute for Complex Systems, National Research Council (ISC-CNR), Uos Sapienza, Piazzale Aldo Moro 5, 00185 Rome, Italy}
\affiliation{Department of Physics, Sapienza Universit\`a di Roma, Piazzale Aldo Moro 1, 00185 Rome, Italy}
\author{Anna Stradner}\email[]{anna.stradner@fkem1.lu.se}
\affiliation{Physical Chemistry, Department of Chemistry, Lund University, SE-221 00 Lund, Sweden}

\begin{abstract}
Concentrated solutions of monoclonal antibodies have attracted considerable attention due to their importance in pharmaceutical formulations, yet their tendency to aggregate and the resulting high solution viscosity has posed considerable problems. It remains a very difficult task to understand and predict the phase behavior and stability of such solutions. Here we present a systematic study of the concentration dependence of the structural and dynamic properties of monoclonal antibodies using a combination of different scattering methods and microrheological experiments. To interpret these data, we use a colloid-inspired approach based on a simple patchy model, which explicitly takes into account the anisotropic shape  and the charge distribution of the molecules. Combining theory, simulations and experiments, we are able to disentangle self-assembly and intermolecular interactions and to quantitatively describe the concentration dependence of structural and dynamic quantities such as the osmotic compressibility, the collective diffusion coefficient and the zero shear viscosity over the entire range of investigated concentrations. This simple patchy model not only allows us to consistently describe the thermodynamic and dynamic behavior of mAb solutions, but also provides a robust estimate of the attraction between their binding sites. It will thus be an ideal starting point for future work on antibody formulations, as it provides a quantitative assessment of the effects of additional excipients or chemical modifications on antibody interactions, and a prediction of their effect on solution viscosity.
\end{abstract}

\maketitle
Immunoglobulin gamma (IgG) constitutes the major antibody isotype found in serum and takes part in the immune response following an infection to the body. IgGs contain three structured domains: two antigen binding domains (FAB) and one so-called constant domain (FC) arranged in a Y shape via a flexible hinge region. The specific details of such a hinge region further classify the IgGs into four subclasses: IgG1, IgG2, IgG3 and IgG4. In the biopharmaceutical industry, monoclonal antibodies (mAb) based on IgGs are a major platform for potential drug candidates, with more than 20 mAb based drugs available on the market and more in development~\cite{Nelson2010, Reichert2012}. The popularity of these macromolecules is due to a large flexibility in molecular recognition thanks to the variable portions of the FAB, a long half-life time in the body, and the possibility of humanization minimizing the risk of immunogenicity.

In order for mAbs to become a successful pharmaceutical product, not only a biological effect but also a high chemical and formulation stability of the solutions is required. Generally, for mAb based drugs, a high concentration formulation of the order of 100 g/L or more is desirable~\cite{Narasimhan2012, Shire2009}. However, in many cases mAb solutions at these concentrations exhibit dramatically altered flow properties, resulting in serious challenges during production and when administering the drug.

The flow properties of protein solutions are primarily determined by the shape of the proteins and their mutual interactions. As the concentration increases, protein-protein interactions become increasingly significant. Despite the extensive experimental and theoretical work devoted to protein crowding and its effects on the resulting stability and flow properties at high protein concentration, our ability to predict for example the concentration dependence of the zero shear viscosity $\eta_0$ and the location of an arrest or glass transition is still limited~\cite{Neergaard2013,Buck2014, Godfrin2016, Grimaldo2014, Yearley2014, Ando2010, Bucciarelli2015, Bucciarelli2016, Foffi2014, Cardinaux2011}. For antibody solutions this is a particularly difficult problem as attractive interactions often lead to reversible self-association between the antibody molecules~\cite{Chari2009, Kanai2008, Yadav2010, Yearley2014, Godfrin2016}, making the change in solution flow properties highly sensitive to the protein concentration~\cite{Lilyestrom2013, Scherer2010, Connolly2012, Schmit2014}.

A number of studies have made attempts to characterize cluster formation in mAb solutions, and to interpret antibody solution properties through analogies with colloids or polymers. In particular, scattering techniques were used to investigate protein interactions and self-association in antibody formulations~\cite{Yadav2012, Yearley2013, Yearley2014, Saito2012, Scherer2013, Castellanos2014, Godfrin2016, Corbett2017}. While investigations of the self-association behavior of various mAb formulations have frequently addressed mAb self-association and its effect on flow properties, we are far from having any predictive understanding and a generally accepted methodology and/or theoretical framework to detect antibody association and model mAb interactions quantitatively. A particular difficulty here is that while the non-spherical shape and internal flexibility has sometimes been addressed, interactions between proteins are frequently treated based on spherical approximations, and in particular the enormous effect that specific, directional interactions can have are generally not considered. 

Here we present an investigation of the solution behavior of a monoclonal antibody varying the concentration, where we combine scattering methods and viscosity measurements with theoretical calculations and Monte Carlo (MC) simulations. We explicitly consider in our model the anisotropy of both the shape and the interactions of the antibody molecules. To this aim we focus on Y-shaped molecules interacting within a simple patchy model that is built from calculations of the electrostatic properties of the considered mAbs.  The simplicity of the model allows for analytical treatment through Wertheim theory \cite{Wertheim1984}, yielding all thermodynamic properties of the solution and in particular the compressibility that can be directly compared to the experimentally determined osmotic compressibility or apparent molecular weight. In addition, we calculate the size distribution of mAb clusters using the Hyperbranched Polymer Theory (HPT) \cite{Rubinstein2003}, without introducing any additional free parameters. Finally, we use MC simulations to verify the results predicted theoretically. 
With the explicit cluster size distribution obtained by HPT at all concentrations investigated, and assuming that the dynamic solution properties (such as the apparent hydrodynamic radius $R_{h,app}$ or the relative viscosity $\eta_r = \eta_0 / \eta_s$, where $\eta_0$ is the zero shear viscosity and $\eta_s$ is the solvent viscosity) are primarily determined by excluded volume effects, we are able to make an additional coarse-graining step in which we model the mAb clusters as effective hard (HS) or sticky (or adhesive) hard (SHS) spheres, for which quantitative relationships for the concentration dependence of $R_{h,app}$ and $\eta_r $ exist. We find that the measured data are indeed well reproduced by this model, confirming that  excluded volume interactions between the assembled clusters are at the origin of the strong increase of $\eta_r $ with increasing concentration.  Hence, our simple model is capable of quantitatively predicting the measured concentration-dependence of the viscosity, solely based on static and dynamic light scattering experiments.  Our results can be easily generalized to different types of mAbs, salt concentrations and temperature and may provide a crucial step for a proper description of self-association and dynamics of monoclonal antibodies.

\section*{Experimental Results}

\begin{figure}
	\centering
	\includegraphics[width=0.95\linewidth]{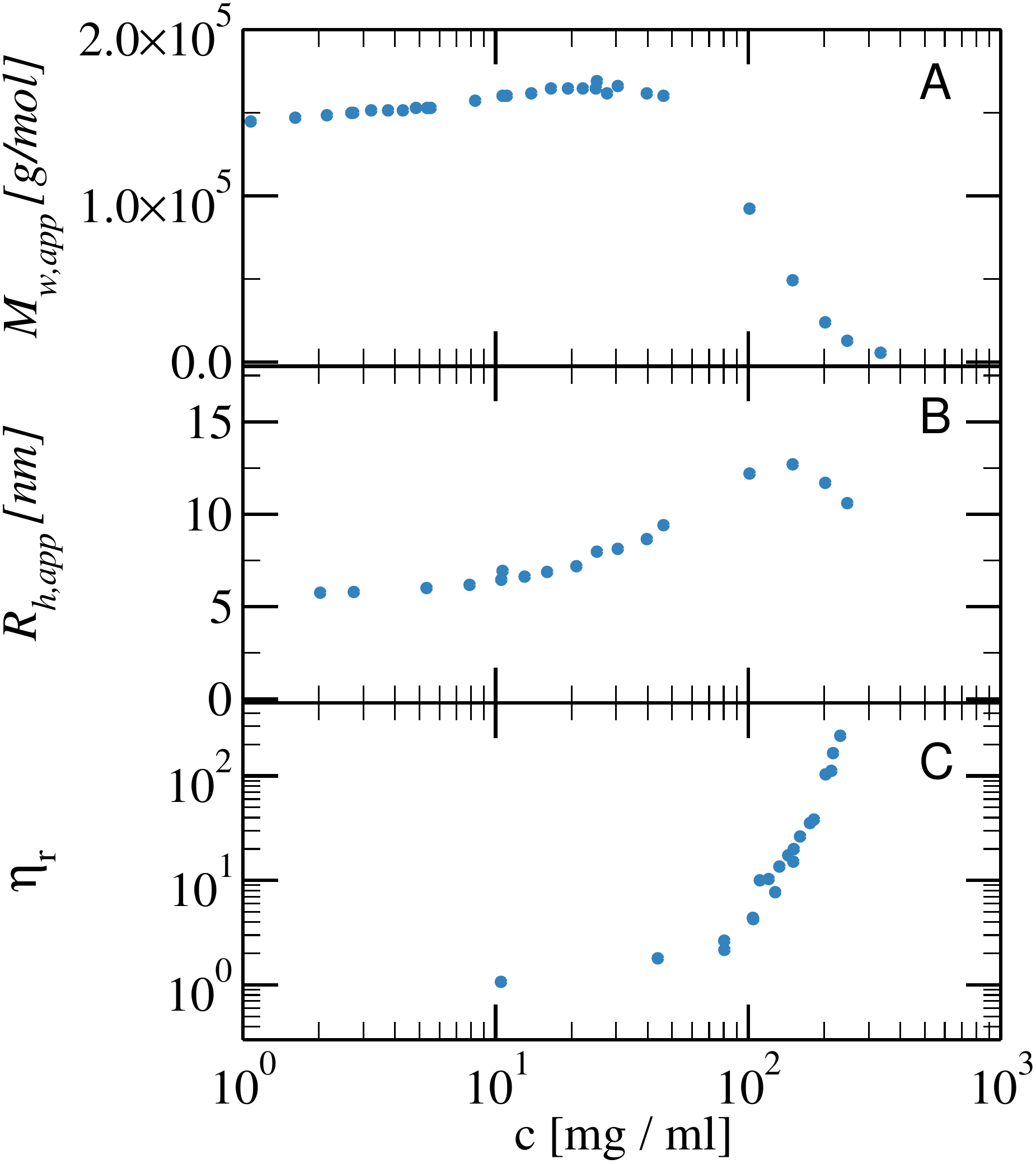}
	\caption{{\bf Experimental results for the concentration dependence of the mAb solutions.} A) Apparent molecular weight $M_{w,app}$ \emph{vs.} weight concentration $C$ as determined by static light scattering. B) Apparent hydrodynamic radius $R_{h,app}$ \emph{vs.} weight concentration $C$ from dynamic light scattering. C) Relative viscosity $\eta_r $ \emph{vs.} weight concentration $C$ measured by DLS-based microrheology.}
	\label{fig:exp-results}
\end{figure}

We have characterized the solution behavior of a monoclonal antibody (mAb) as described in \emph{Materials and Methods}. The results from these experiments are summarized in Fig.~\ref{fig:exp-results}. The static light scattering (SLS) data in Fig.~\ref{fig:exp-results}A show that the apparent molecular weight $M_{w,app}$ initially increases with concentration $C$ from the known value of the molecular weight of the mAb monomer, i.e. $M_1 = 147000$ g/mol, goes through a maximum at a concentration of around $C \approx 30$ mg/ml, and then strongly decreases at higher concentrations. A similar trend can also be seen for the apparent hydrodynamic radius $R_{h,app}$, reported in Fig. ~\ref{fig:exp-results}B, that is obtained by dynamic light scattering (DLS). We find that  also $R_{h,app}$ initially increases from the monomer value of $R_{h,app} \approx 6$ nm, reaches a maximum at $C \approx 150$ mg/ml, and finally decreases at higher values of $C$. In contrast, the reduced viscosity $\eta_r$, shown in Fig. ~\ref{fig:exp-results}C, monotonically increases with concentration and appears to diverge for $C \approx 200 - 300$ mg/ml. Qualitatively, the concentration dependence of the three key quantities $M_{w,app}$, $R_{h,app}$ and $\eta_r$ is in agreement with a behavior where the mAb self-assemble into aggregates with increasing concentration. While this is visible in the SLS and DLS data at low concentrations, the influence of excluded volume effects on the scattering data becomes more prominent at higher concentrations and results in a decrease of the measured values for $M_{w,app}$ and $R_{h,app}$. At the same time, these increasing interaction effects also result in a corresponding increase of the zero shear viscosity of the mAb solution.

While it is straightforward to qualitatively assess the existence of aggregation and intermolecular interactions, a quantitative interpretation of the experimental data would require knowledge of both the molecular weight distribution of the resulting aggregates as well as the interaction potential between antibodies. This situation is similar to the difficulties encountered when trying to analyze scattering and rheology data of surfactant molecules forming large polymer-like micelles \cite{Schurtenberger1989, Schurtenberger1996}. Crucially, a qualitative comparison between  the behavior normally encountered for polymer-like micelles and the data shown in Fig. \ref{fig:exp-results}  shows significant differences. Indeed, for polymer-like micelles the maxima in $M_{w,app}$ and $R_{h,app}$ are directly linked to the overlap concentration $C^*$ that marks the transition from a dilute to a semi-dilute concentration regime, and thus occur at approximately the same value. For the mAb data shown in Fig. \ref{fig:exp-results}, however, there exists a large difference between the concentrations related to the maxima in $M_{w,app}$ and $R_{h,app}$, respectively. This clearly indicates that a simple application of polymer models, such as the wormlike chain model previously used successfully to for example describe SLS and DLS data for antigen-mAb complexes \cite{Murphy1988}, does not work. We thus instead exploit analogies to patchy colloids in order to design a coarse-grained model for our system and investigate whether we can obtain with this approach a quantitative analysis of the experimental data.

\begin{figure*}
	\centering
	\includegraphics[width=1\linewidth]{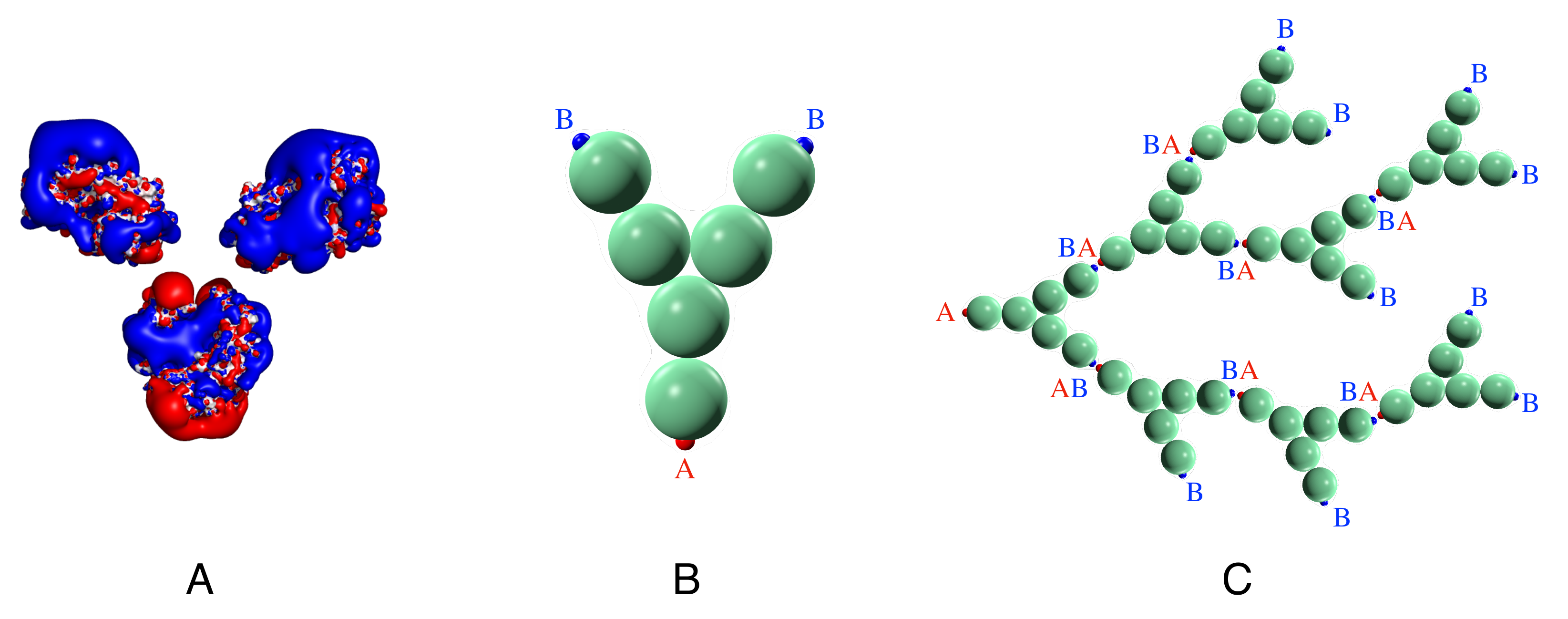}
	\caption{{\bf Design of the patchy model of mAbs condensing electrostatic interactions into patch-patch specific attraction.} A: isosurfaces of the -1 (red) and +1 $k_BT$ (blue) electrostatic potential at pH $6.5$ with $10$ $mM$ NaCl, indicating an overall positive charge for the arms (FAB domains) and a largely negative charge for the tail (FC domain); B: illustration of the patchy YAB model: 6 hard spheres (in green) each of diameter $\sigma$ are constrained to a rigid Y shape, constituting a single mAb molecule. Each molecule is decorated with one $A$ patch on the tail (red) and two $B$ (blue) patches, one on each arm, respectively. Only $AB$ attractive interactions are considered mimicking the arm-to-tail electrostatic interactions; C: schematic view of the clustering process for mAb molecules forming hyperbranched structures.
	}
	\label{fig:model}
\end{figure*}

\section*{Comparing theory and experimental results}
\subsection*{Model: Antibodies as patchy particles}

We model mAbs as patchy colloids and use a theoretical approach that has previously been applied successfully to such particles, in order to calculate their structural properties as a function of concentration. Patchy models are coarse-grained models which condense complex anisotropic interactions often of electrostatic origin in simple site-site aggregation, that have been applied in the past to several protein solutions\cite{fusco2014characterizing,roosen2014ion,li2015charge,quinn2015fluorescent,mcmanus2016physics,cai2017eye,cai2018proof}, and other complex systems, including colloidal clays\cite{RuzickaNatMat} and DNA-based nanoconstructs\cite{biffi2015equilibrium,bomboi2016re}. 

In order to build a meaningful model it is crucial to identify the key ingredients controlling the intermolecular interactions. A previous study of this antibody has shown that the viscosity is sensitive to the salt concentration, pointing towards electrostatic interactions as a main component of the intermolecular interactions\cite{Neergaard2013}. Therefore,  we first carry out a study of the electrostatic isosurface of a single antibody molecule in the considered buffer solution, as described in \emph{Materials and Methods}, in order to locate the active spots on the molecule surface that are involved in particle-particle aggregation. The resulting charge distribution is illustrated in Fig.~\ref{fig:model}A, which clearly shows that the considered mAbs have an overall positively charged
surface on the two arms (FAB domains) and a largely negative charge on the tail (FC domain). This suggests that the main driving mechanism for mAbs aggregation has to be an attractive arm-to-tail interaction.

To take into account this result, we thus consider Y-shaped particles formed by six spheres of diameter $\sigma$ and decorated with three patches, one of type $A$ on the tail and two of type $B$ on the arms, as illustrated in Fig.~\ref{fig:model}B. Interactions between $AB$ patches are attractive and modeled with a square-well potential, while $AA$ and $BB$ interactions are not considered. To predict the behavior of our patchy model, which we call YAB model,  we use a thermodynamic perturbation theory, introduced by Wertheim roughly 30 years ago, which describes associating molecules under the hypothesis that each sticky site on a particle cannot bind simultaneously to two or more sites on another particle \cite{Wertheim1984}.
The Helmholtz free energy and the thermodynamic properties of the system, including for example the energy per particle, the specific heat at constant volume and the isothermal compressibility, can thus be predicted from the dependence of the bonding probability $p$ on the temperature $T$ and the number density $\rho$, as explained in more details in \emph{Materials and Methods}. We complement this approach with Monte Carlo simulations of the YAB model in order to validate the theoretical results. 
In addition, the YAB model belongs to the class of hyperbranched polymers\cite{Rubinstein2003}, for which it is possible to calculate the equilibrium cluster size distribution of the clusters solely from the knowledge of the bonding probability $p$ (see \emph{Materials and Methods}). As this parameter is directly an outcome of Wertheim theory, the YAB model is amenable to a full analytical treatment, allowing one to obtain simultaneously the thermodynamic and the connectivity properties of the solutions, to be directly compared with the experimental results.

\subsection*{Comparison between theory and MC simulations}
\begin{figure*}
	\centering
	\includegraphics[width=0.9\linewidth]{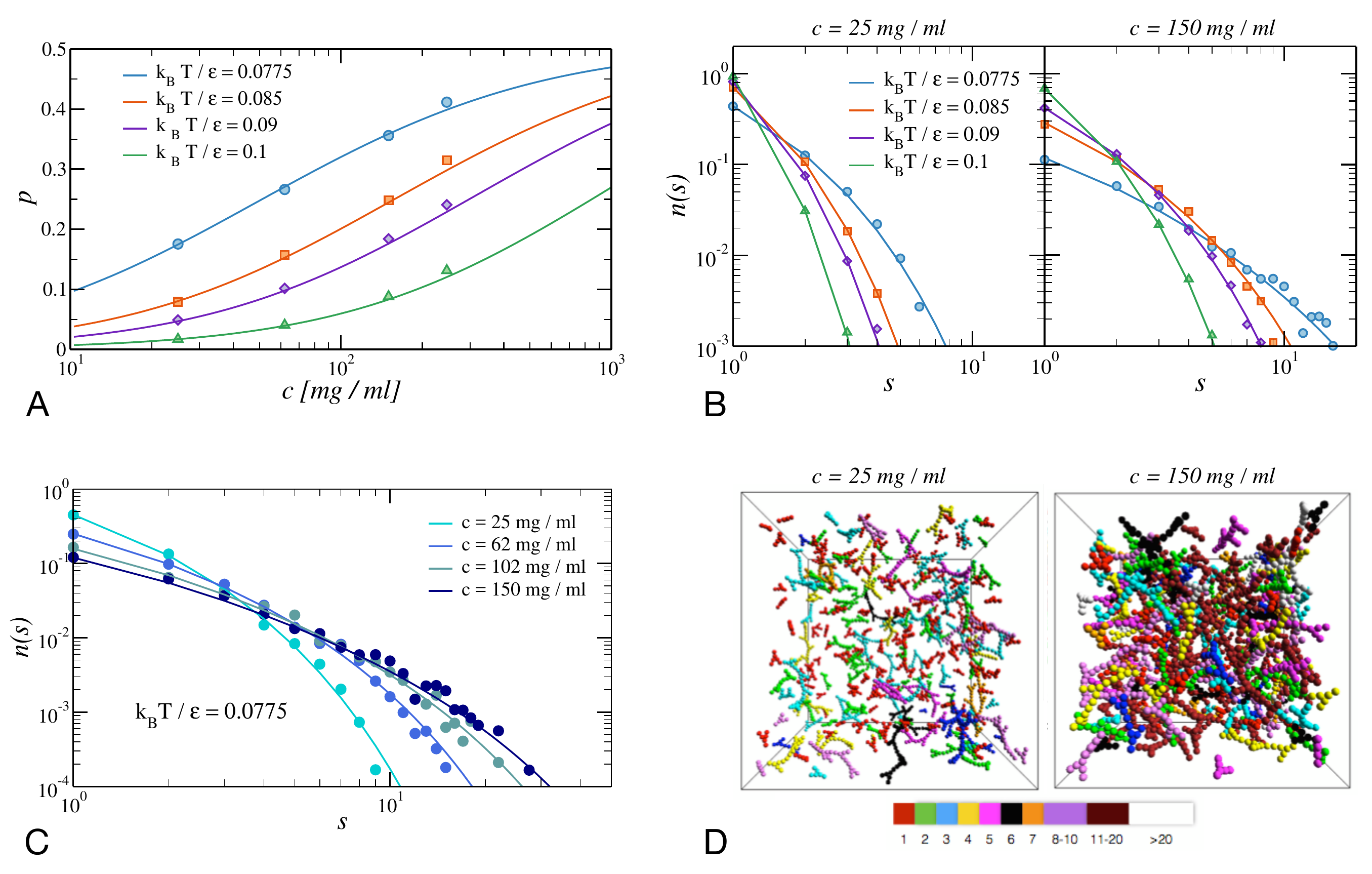}
	\caption{{\bf Results from theory (lines) and simulations (symbols).} A: Bond probability $p$ as a function of mAb concentration at different attraction strengths $k_BT/\epsilon$. Small deviations between theory and simulations only occur at high $C$; B: Cluster size distributions $n(s)$ for $c = 25$ $mg/ml$ (left) and $c = 150$ $mg/ml$ (right) for different $k_BT/\epsilon$; C: Cluster size distributions $n(s)$ for $k_BT/\epsilon=0.0075$ and several values of $C$; D: Simulation snapshots $c = 25$ $mg/ml$ (left) and $c = 150$ $mg/ml$ (right). Different colours identify different cluster sizes.}	
	\label{fig:snp-cl}
\end{figure*}
The mAbs modeled as patchy Y-shaped colloids self-associate into clusters with increasing concentration through reversible $AB$ bonds, as a result from the attraction between $A$ and $B$ patches. Their assembly can be monitored by focusing on the variation of the bonding probability $p$ and the distribution $n(s)$ of clusters of size $s$ as a function of the two parameters controlling the assembly: the attractive strength $k_B T / \epsilon$, where $\epsilon$ is the well depth of the square-well attraction between $A$ and $B$ patches (see \emph{Materials and Methods}), $T$ is the temperature and $k_B$ is the Boltzmann constant, and the mAb concentration $C$. 

We report in Fig.~\ref{fig:snp-cl} some representative results comparing theory and simulations, including the bond probability and the cluster size distributions for different concentrations and attraction strengths. In all cases, we find that there is quantitative agreement between theory and simulations for both thermodynamics and cluster observables. Thus, we can confidently use the results of the theoretical approach in order to compare with experimental results.

\subsection*{Structural Properties}
In order to analyze the measured $M_{w,app}$, we calculate the isothermal compressibility $\kappa_T=-1/V (\partial V/\partial P)_T$ for our YAB model, since $\kappa_T$ is related to the
$S(0)$, the static structure factor at $q = 0$, as
\begin{equation}\label{S0}
	S(0)=\rho k_B T\kappa_T,
\end{equation}
\noindent which in turn is related to the experimentally determined apparent weight average molar mass by $M_{w,app} = M_{1}S(0)$ where $M_1$ is the molar mass of a monomer. 
In a solution where antibodies self-assemble into larger clusters described via Wertheim theory, static light scattering thus provides an apparent weight average aggregation number $N_{app}$ given simply by

\begin{equation}\label{naggapp}
	N_{app} = S(0),
\end{equation}
\noindent where $N_{app} = M_{w,app}/M_1$ is the apparent aggregation number, with $M_1$ being the molar mass of a monomer.

When trying to understand self-assembly in mAb solutions, we need to be able to account for both the average aggregation number, $N_{agg}$, as well as the resulting interaction effects between the antibody clusters, given by $S(0)$. 
Using Wertheim theory, we can calculate the free energy and differentiate it twice in order to get $\kappa_T$. As described in more details in \emph{Materials and Methods}, the free energy is the sum of a hard-sphere reference term plus a bonding term. The reference HS term is the Carnahan-Starling (CS) free energy of an equivalent HS system. Since mAbs are not spherical, we cannot directly use the actual volume fraction given by the number density of mAbs and the volume of a monomer, but we rather need to determine an equivalent hard sphere diameter $\sigma_{HS}$ of the Y-molecule.
We thus calculate $\kappa_T$ for different values of $\sigma_{HS}$ and $k_B T / \epsilon$ and compare it to the measured data.

\begin{figure}[h]
\centering
		\includegraphics[width=0.9\linewidth]{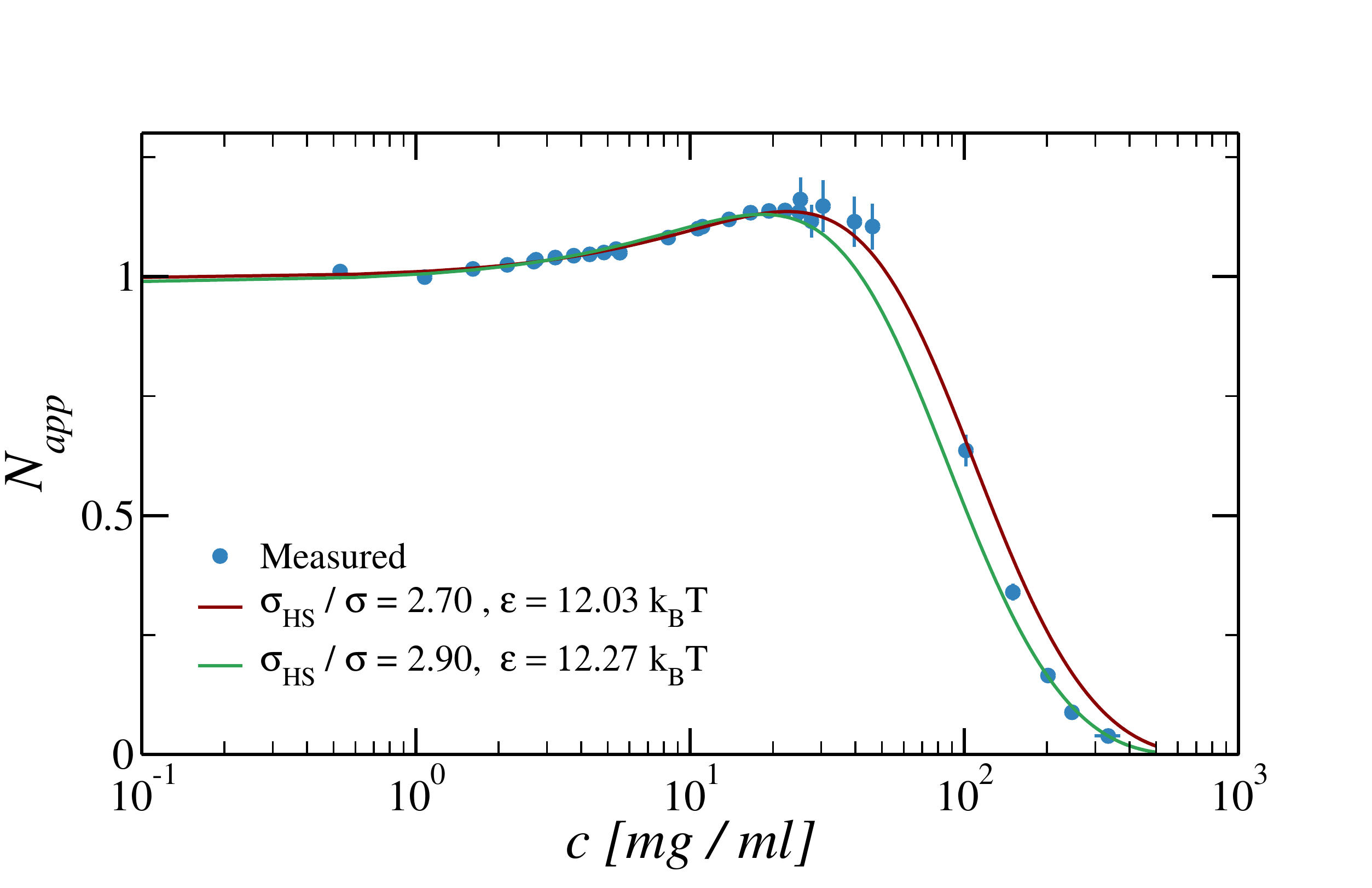}
	\caption{{\bf Comparison of SLS data with patchy model predictions.} Experimental $N_{app}$ compared with YAB model results: the best agreement, particularly for high concentration data is obtained for an equivalent hard sphere diameter $\sigma_{HS}=2.90\sigma\sim4.2$nm  and $\epsilon/k_BT=12.27$.
	}
	\label{fig:S0}
\end{figure}
\begin{figure*}[t]
	\centering\includegraphics[width=0.7\linewidth]{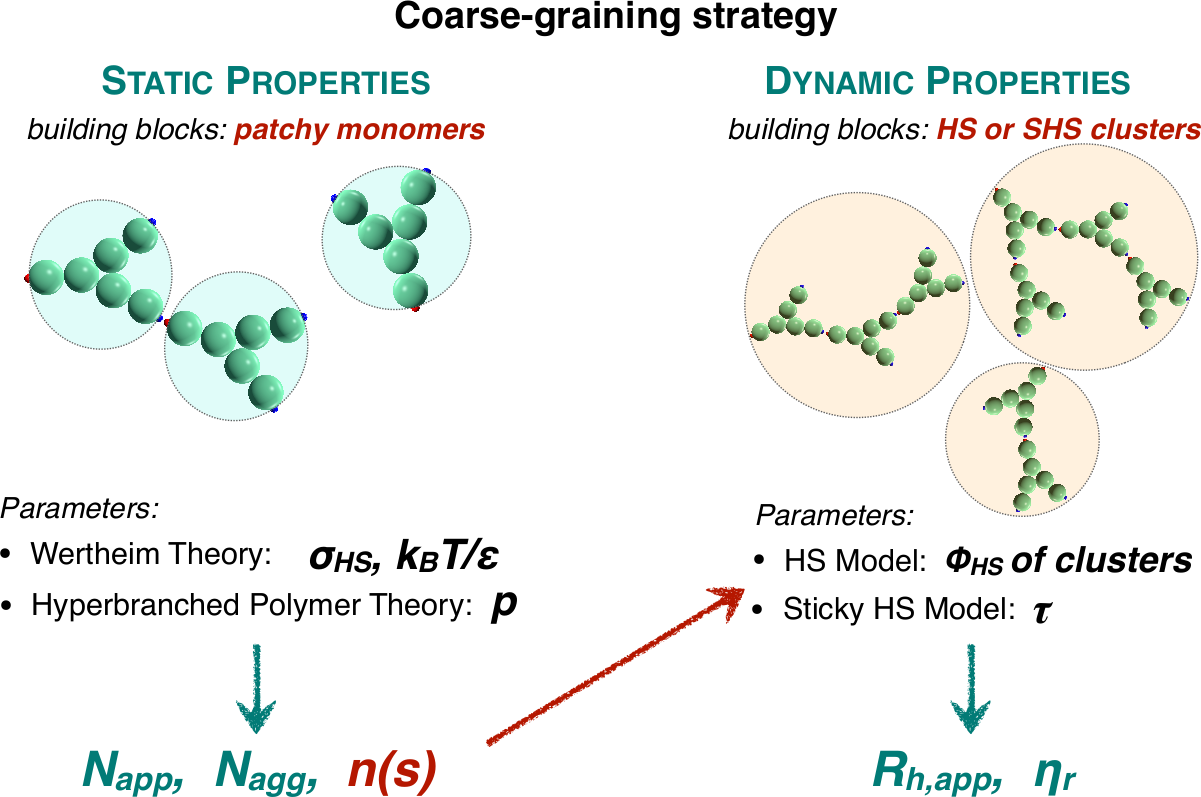}
	\caption{{\bf Schematic illustration of the coarse-graining strategy} used to analyze the concentration-dependent cluster formation and its effects on the structural and dynamic properties of the solution.}
	\label{fig:strategy}
\end{figure*}
By fitting the theoretical results to the experiments as described in \emph{Materials and Methods}, we determine the two unknown parameters: the strength of the $AB$ interaction and the equivalent HS diameter. Fig.~\ref{fig:S0} compares $N_{app}$ for the YAB model to the SLS data
and we find that the best fit of the data, particularly correctly describing the high concentration behavior which is most relevant for the viscosity to be discussed later,  is obtained with an effective hard sphere diameter of $\sigma_{HS} = 2.9\sigma$  and a strength of the AB patch-patch attraction given by $\epsilon \simeq 12.3 k_BT$. 
Note that the estimated value of $\sigma_{HS}$ is considerably smaller than the geometric diameter of the Y molecule, thus accounting for the penetrability of the Y-shaped antibodies. When converted in real units, an effective HS radius of $4.2$ $nm$ is found, which also compares well with the measured radius of gyration of the antibody molecule $ R_g \approx 4.7$ $nm$.

\subsection*{Dynamic Properties}
Having analyzed the SLS data using Wertheim theory, we now have a prediction for the effect of concentration on the self-assembling behavior of mAbs and we can thus calculate the cluster size distributions at all concentrations thanks to HPT. Next we make an attempt to test the consistency of these results with the data obtained using DLS for the same samples shown in Fig. \ref{fig:exp-results}B. Unfortunately, this is much less straightforward than the analysis of the SLS data and requires an additional coarse graining step, illustrated in Fig.~\ref{fig:strategy}. The main problem here is that we currently lack a theoretical model that would allow us to calculate the effective or apparent hydrodynamic radius of concentrated solutions of polydisperse antibody clusters. 
We thus propose an approach in which we use the self-assembled clusters of the patchy model and treat them as new interacting objects. Their dominant interaction is of course excluded-volume and, hence, we consider them as effective polydisperse hard spheres, each with its own radius resulting from its size in terms of monomers. To go one more step, we also consider them as sticky hard spheres.

Within this approach we first calculate the $z$-average~\cite{PeterLectures} hydrodynamic radius $R_{h,z}$ of the mAb solutions using the cluster size distributions obtained theoretically. Next we model the solutions at each concentration as dispersions of colloids with a size given by $R_{h,z}$ and an effective hard sphere volume fraction $\phi_{HS}$. The influence of interparticle interactions on the resulting collective diffusion coefficient, or $R_{h,app}$, is calculated by treating the spheres either as hard or sticky hard spheres, for which accurate expressions exist.

First, we need to determine the hydrodynamic radius $R_h$ of mAb clusters of a given size $N_{agg}$. Clusters of mAbs of a given size $N_{agg}$ were generated randomly, where the clusters also have to satisfy the criterion of self-avoidance and where each monomer in a cluster is allowed to have a maximum of 3 connections, i.e. reflecting the YAB structure imposed in Wertheim theory and HPT. For each individual cluster its hydrodynamic radius was then calculated using the program Hydropro\cite{Ortega2011}, and average values were calculated from 100 individual clusters. This resulted in a data set of $R_h$ \emph{vs} $N_{agg}$ that was well reproduced by the phenomenological relationship $R_h = 3.69 + 2.04 \times N_{agg} - 0.069 \times N_{agg}^2$, where $R_h$ is given in $nm$.

With this relationship and assuming hard sphere-like interactions between the different clusters, we can now calculate the concentration dependence of both $N_{app}$ and $R_{h,app}$. The expression for the measured apparent molecular mass in this coarse grained model is $M_{w,app} = M_{w}S^{eff}(0)$, where $M_{w}$ is the weight average molar mass of the clusters. Note that the static structure factor $S^{eff}$ introduced here has a different definition than $S(0)$ introduced in Eq.~\ref {S0}, and $S^{eff} = S(0)/N_{agg}$ now corresponds to the effective structure factor of a solution of polydisperse spheres, reflecting the fact that the mAb clusters and not the individual antibodies are the new interacting objects. The apparent weight average aggregation number $N_{app}$ is then given by\cite{PeterLectures}
	\begin{equation}\label{naggapp_col}
	N_{app} = N_{agg} S^{eff}(0).
	\end{equation}	
The only adjustable parameter introduced by this step is the conversion of the weight concentration into the effective hard sphere volume fraction $\phi_{HS}$ of the clusters.
For hard spheres, we can exploit the Carnahan-Starling expression for the low wavevector limit of the static structure factor, 
\begin{equation}\label{C-S}
	S_{CS}(0) = \frac{(1 - \phi_{HS})^4}{(1 + 2 \phi_{HS})^2 + \phi_{HS}^3 (\phi_{HS} - 4)},
\end{equation}
as well as the weight average aggregation number $N_{agg}$, obtained with Wertheim theory and HPT, in order to calculate $N_{app}$ using Eq.~\ref {naggapp_col}.  In doing these calculations we fix the effective diameter $\sigma_{HS} = 2.9 \sigma$ of each antibody molecule (Fig.~\ref{fig:S0}). 
\begin{figure}[h]
	\centering
	\includegraphics[width=0.9\linewidth]{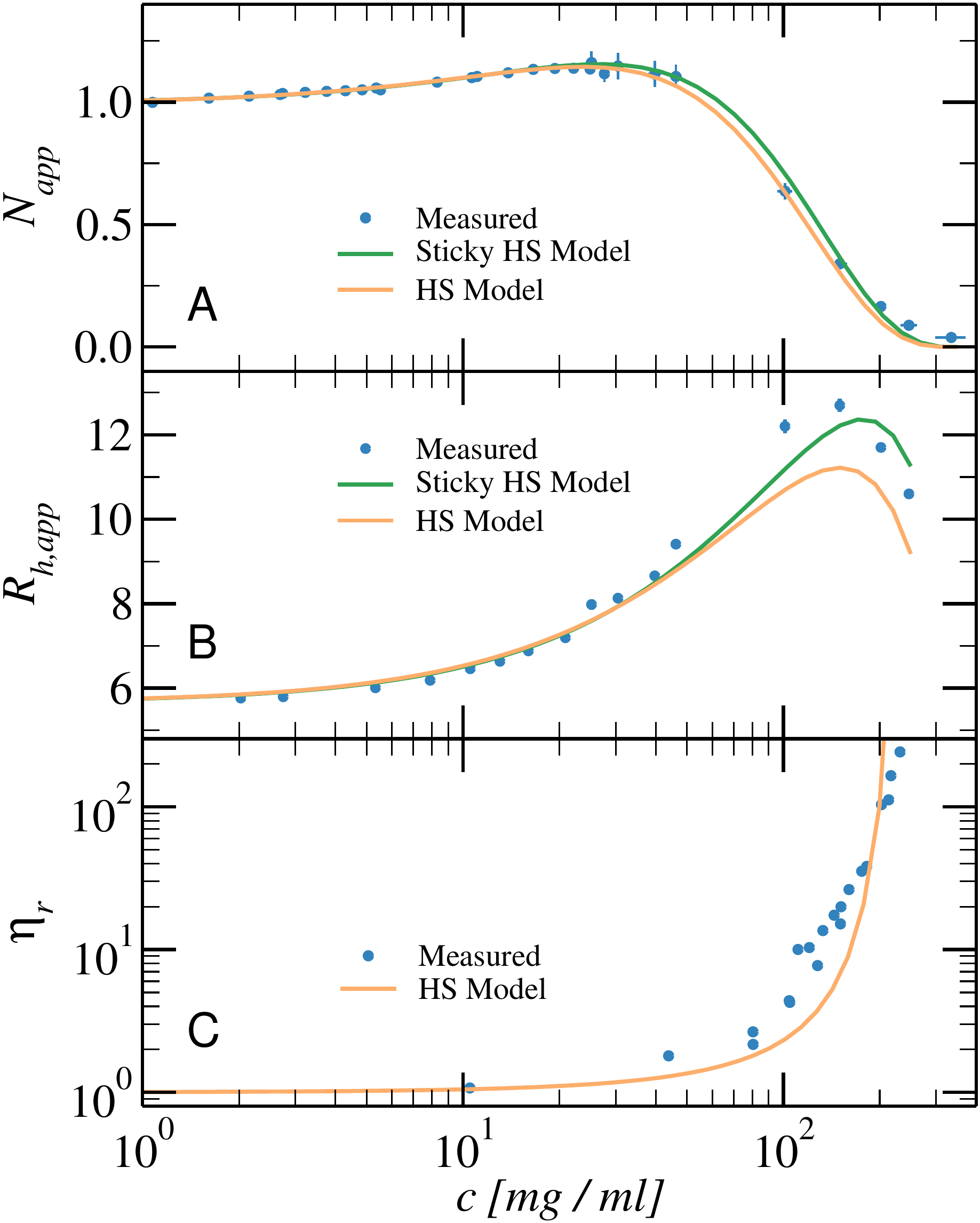}
	\caption{{\bf Comparison between experimental and theoretical results for the concentration dependence of static and dynamic properties of the mAb solutions.} Blue symbols are experimental data, while solid lines are the theoretical data for the hard sphere (orange line) and the sticky hard sphere (green line) models, respectively. The fit parameters are reported in Table\protect\ref{tab:fitresults}. A: Apparent aggregation number $N_{app}$ versus weight concentration as determined by SLS;  B: Apparent hydrodynamic radius $R_{h,app}$ versus weight concentration from dynamic light scattering; C: Reduced viscosity $\eta_r$ versus weight concentration measured by DLS-based microrheology.}
	\label{fig:final}
\end{figure}

The effective cluster HS volume fraction is calculated taking into account that the excluded volume contribution of an antibody in a cluster is equal to a sphere with a radius equal to the antibody radius of gyration and also that clusters are fractal, giving 
\begin{equation}\label{phiHS}
\phi_{HS} = \left(\frac{2 R_g}{\sigma_{HS}}\right)^3\!\! \phi \ N_{agg}^{(3 - d_F)/d_F} =1.41\phi \ N_{agg}^{(3 - d_F)/d_F},
\end{equation} 
where $d_F = 2.5$ is the fractal dimension of the clusters and $\phi$ is the nominal antibody volume fraction ($\phi=\pi/6\rho d^3$) based on the geometric diameter $d$ of the molecule. Thus, in the coarse grained model we have an effective hard sphere volume fraction that is $\approx40$\% higher than for the individual mAbs in the Wertheim analysis, which does not seem unrealistic because clusters cannot overlap as much as individual antibodies do. The resulting comparison of the model calculations  with experiments provides a very good description of the data, as shown in Fig.~\ref{fig:final} A.

In order to calculate $R_{h,app}$ we use the corresponding virial expression for the short time collective diffusion coefficient, which results in
\begin{equation}
	\label{DHS}
	R_{h,app} = R_{h} / (1 + k_D \phi_{HS}),
\end{equation}
\noindent where $k_D = 1.45$ for hard spheres \cite{Banchio2008}. Note that here we use the $z$-average aggregation number in order to calculate $R_{h}$. The agreement for $R_{h,app}$  with the results from the simple hard sphere model is quite good  (Fig.~\ref{fig:final}B), except for the highest values of $C$, where we expect Eq.~\ref{DHS} to fail and would instead need to include higher order terms. We also find that the apparent hydrodynamic radius obtained in DLS experiments is very sensitive to the interparticle interactions, and we can thus also look at a somewhat refined interaction model, where we also include the possibility of an additional weak attraction between different clusters. Here we use the so-called adhesive or sticky hard sphere model \cite{Piazza1998, Cichocki1990}, where we include an additional weak short-range attractive potential that could be due to the unbound attractive patches of the mAbs at the exterior of the clusters. In this model, Eqs.~\ref{C-S} and ~\ref{DHS} then become
\begin{equation}
	\label{S(0)-stsph}
	S_{SHS}(0) = \frac{(1 - \phi_{HS})^4}{(1 + 2 \phi_{HS} - \lambda \phi_{HS})^2},
\end{equation}
and 
\begin{equation}
	\label{DStSph}
	R_{h,app} = R_{h} / (1 + (1.45 - 1.125/\tau) \phi_{HS}),
\end{equation}

\noindent where $\tau$ is the stickiness parameter that is inversely proportional to the strength of the attractive interaction and $\lambda$ is given by
\begin{equation}\label{lambda}
	\lambda = 6 (1 - \tau + \tau / \phi) \biggl(1-\sqrt{1 - \frac{1 + 2 / \phi}{6 (1 - \tau + \tau / \phi)^2}}\biggr).
\end{equation}
The corresponding theoretical curves when $\tau$ is used as an additional fit parameter to the SLS and DLS data are also shown in Fig.~\ref{fig:final}A and B. In particular, a better description of the apparent hydrodynamic radius is obtained within the SHS model with
 $\tau\sim 2.5$, corresponding to a very weak additional attraction between the mAb clusters. While the approximations made in our coarse grained strategy may be too severe to say much about the exact nature of the effective interaction potential between the mAb clusters in solution, the experimental data are very well reproduced by our simple model. This indicates that the two chosen models, a pure hard sphere and an adhesive hard sphere interaction with moderate stickiness, likely bracket the true behavior of the self-assembling antibody investigated in this study.

Finally, as an ultimate test, we calculate the concentration dependence of the relative viscosity $\eta_r$. We use the expression for $\eta_r$ developed by Mooney, which is often and successfully applied for mono- and polydisperse hard sphere colloidal suspensions \cite{Mooney1951}:
\begin{equation}\label{Mooney}
	\eta_r = e^{\frac{A\phi_{HS}}{(1 - \phi_{HS}/\phi_g)}}.
\end{equation}
Here $A$ is a constant, which for hard spheres is 2.5, and $\phi_g$ is the maximum packing fraction, which depends on the polydispersity of the system. In order to estimate it, we have evaluated the polydispersity of our antibody clusters as a function of concentration and find that at the highest studied concentration it reaches about 45\%. For such polydisperse hard spheres, the maximum packing fraction is $\approx 0.71$\cite{Farr2009}. Using this value, we then should directly obtain the concentration dependence of $\eta_r$ from the previously determined relationship between $C$ and $\phi_{HS}$ without any free parameter. The resulting comparison between the measured and calculated values of $\eta_r$ is shown in the bottom panel of Fig. \ref{fig:final}, and the agreement is indeed quite remarkable given the lack of any free parameter. This clearly indicates that it is the excluded volume interactions between the self-associating clusters that is at the origin of the strong increase of the zero shear viscosity with increasing concentration, and our simple model is capable of predicting the measured $C$-dependence based on static and dynamic light scattering experiments quantitatively.

\section*{Discussion and Conclusions}
The self-assembly of monoclonal antibodies and its effect on the solution properties such as the viscosity is an important factor in determining our ability to develop high concentration formulations. However, there has been a lack of decisive experimental and theoretical approaches to obtain a quantitative and predictive understanding of antibody solutions. 
A recent theoretical study has proposed a patchy model for antibody molecules\cite{kastelic2017controlling}, in which different types of patch-patch attractions were considered resulting in a large number of parameters to be adjusted to describe different experimental conditions. On the other hand, in the present work we define the simplest model based on electrostatic calculations for the specific type of immunoglobulin also studied experimentally within the same buffer and salt conditions. This very simple model is analytically solvable by well-established theories, in particular the combination of the Wertheim theory with hyperbranched polymer theory to predict the aggregation properties of the mAb solutions. We have also shown that both thermodynamic properties and cluster distributions are in quantitative agreement with MC simulations of the model, thus the theoretical predictions can be directly compared with experiments without suffering from numerical uncertainty. 
From the mAb self-assembly process built by the patchy interactions, we then employ a second coarse-graining step in which we consider our antibody clusters as the elementary units. We thus use the most basic description considering these clusters interacting essentially as hard spheres or sticky hard spheres with very moderate attraction, and apply available phenomenological descriptions to predict the dynamic properties of the system. This treatment does essentially not depend on any free parameter and is able to reproduce all measured data from SLS, DLS and microrheology. This simple model, based on very fundamental assumptions, thus provides an elegant way to consistently describe the thermodynamic and dynamical behavior of mAb solutions. 

The patchy model that we have established also provides a robust estimate of the attraction between patchy binding sites through Wertheim theory, and thus will be an ideal starting point to investigate and quantitatively assess the effects of additional excipients or chemical modifications on the antibody interaction. Such information is vital for an advanced formulation strategy and attempts to predict antibody stability and the resulting viscosity from molecular information. Moreover, the combination of static scattering data and Wertheim/HPT to determine the interaction strength and the cluster size distribution as a function of concentration, and the subsequent test using DLS and (micro)rheology measurements without additional free parameters other than a rescaling of the volume fraction, allows us to critically test models for the type of interactions responsible for the self-association of a given mAb into clusters.

\section{Materials and Methods}
\subsection*{Sample preparation}
 The mAb used in this study was a humanized IgG4 against trinitrophenyl , which was previously found to exhibit an increased viscosity at high concentrations \cite{Neergaard2013} (where it was labeled \textit{mab-C}). It was manufactured by Novo Nordisk A/S and purified using Protein A chromatography, and subsequently concentrated to 100 mg/ml and buffer exchanged into a 10 mM Histidine buffer with 10 mM NaCl at pH 6.5. 
	
	For measurements, the sample was diluted and buffer exchanged to a 20mM Histidine pH 6.5 buffer containing 10 mM NaCl and subsequently concentrated using a 100 kD cutoff spinfilter (Amicon inc.). The concentrated sample was used as a stock solution for preparing the less concentrated ones. The concentration of each sample was determined by a series of dilutions followed by measurement of the absorption at 280nm using an extinction coefficient of $e^{280nm}_{1\%,1cm} = 2.234$.  In order to assess the uncertainty of the concentration determination the dilution series was done in triplicates.
	
	\subsection*{Light Scattering}
	The dynamic and static light scattering experiments were made using a 3D-LS Spectrometer (LS Instruments AG, Switzerland) with a 632nm laser, recording DLS and SLS data simultaneously. The measurements were conducted at $90^{\circ}$ scattering angle. Before measurement, the samples were transferred to pre-cleaned 5mm NMR tubes  and centrifuged at 3000 g and 25 $^{\circ}$C for 15 min, to remove any large particles and to equilibrate temperature. Directly after centrifugation, the samples were placed in the temperature equilibrated sample vat and the measurement was started after 5 minutes to allow for thermal equilibration. Additional low concentration SLS measurements were done using a HELIOS DAWN multi-angle light scattering instrument (Wyatt Technology Corporation, CA, USA), connected to a concentration gradient pump. Both instruments were calibrated to absolute scale using a secondary standard, allowing  for direct comparison of the two data sets.

\subsection*{Microrheology}
	The zero shear viscosity $\eta_0$ was obtained using DLS-based tracer microrheology. Sterically stabilized (pegylated) latex particles were mixed with protein samples to a concentration of 0.01 $\% $v/v using vortexing and transferred to 5 mm NMR tubes.
	The sterically stabilized particles were prepared by cross-linking 0.75 kDa amine-PEG (poly-ethylene glycol) (Rapp Polymere, 12750-2) to carboxylate stabilized polystyrene (PS) particles (ThermoFischer Scientific, C37483) with a diameter of 1.0 $\mu$m using EDC (N-(3-Dimethylaminopropyl)-N'-ethylcarbodiimide) (Sigma Aldrich, 39391) as described in detail in \cite{Garting2018}.
DLS measurements were performed on a 3D-LS Spectrometer (LS Instruments AG, Switzerland) at a scattering angle of 46-50$^\circ$ to stay away from the particle form factor minima and thus to maximise the scattering contribution from the tracer particles with respect to the protein scattering. Measurements were made using modulated 3D cross correlation DLS \cite{Block2010} to suppress all contributions from multiple scattering that occur in the attempt to achieve conditions where the total scattering intensity is dominated by the contribution from the tracer particles. Samples were either prepared individually or diluted from more concentrated samples using a particle dispersion with the same particle concentration as in the sample as the diluent. The diffusion coefficient $D$ of the particles was then extracted from the intensity autocorrelation function using a 1st order cumulant analysis of the relevant decay. This diffusion coefficient is compared to that of particles in a protein-free buffer and the relative viscosity is extracted from the relationship between diffusion coefficient and viscosity in the Stokes-Einstein equation given by $D = k_BT / 6 \pi \eta_0 R_h$, where $R_h$ is the known hydrodynamic radius of the tracer particles\cite{Garting2018, Furst2017}.
	
\subsection*{Isosurface calculations}
The FAB domains were built using the antibody modeler tool in the Molecular Operating Environment (Chemical Computing Group Inc, Canada) computer program\cite{MOE2011}, whereas the FC domain was taken from a crystallographic structure with a similar FC domain found in the protein data bank (PDBID: 4B53)
The electrostatic calculations were done in a two step process, using pdb2prq \cite{Dolinsky2007}  and the automated poisson Boltzmann solver\cite{Jurrus2018} (apbs) pymol plugin. The pdb2pqr server is hosted by the  National Biomedical Computation Resource at http://nbcr-222.ucsd.edu/pdb2pqr\_2.1.1/, and was used to calculate the protonation state of the FAB and FC domains at pH $6.5$ taking the local structure around the titratable residues into account. The prepared structures were then used by the apbs plugin to calculate an electrostatic map of the protein. The apbs was run using the default parameters, with the addition of Na+ and Cl- ions corresponding to a salt concentration of 10mM.
	
\subsection*{YAB Patchy model and MC simulations}
The antibody molecule is represented as a  symmetric $Y$-shaped particle, constructed from six hard spheres of diameter $\sigma$, as illustrated in Fig. \ref{fig:model}B. Each mAb is decorated by 3 patches, one of type $A$ on the tail and two of type $B$ on the arms. Only $AB$ interactions are taken into account based on the charge distribution on the surface of the mAb molecule in the studied buffer conditions, and are modeled as an attractive square well (SW) potential of range $\delta = 0.1197\sigma$, which guarantees that each patch is engaged at most in one bond. For this model the geometric diameter $d$ of a single mAb molecule is that of the circle tangent to the external spheres: $d = \frac{9 + 2 \sqrt{3}}{3} \sigma$.

We perform standard MC simulations of $N=1000$ YAB particles at different number densities $\rho=N/V$ where $V$ is the volume of the cubic simulation box. The unit of length is $\sigma$. To compare the experimental value of $C$ with simulations and theory, we consider the geometric radius $d/2$ of the Y-colloid equal to the hydrodynamic radius measured for a single mAb molecule, that is $\approx 6$nm.  With this choice we have that $\sigma \approx 2.89$nm and, considering that the mass of a molecule is 150 kDa, an experimental concentration of 1 mg/ml corresponds to $9.6938 \times 10^{-5}/\sigma^3$ in simulation units.

\subsection*{Theory}
In Wertheim theory\cite{Wertheim1984,Tavares2010}, the free energy $F$ of a system of $N$ particles in a volume $V$, with number density $\rho=N/V$, is calculated as the sum of a hard sphere reference term plus a bonding term. The bonding free energy $F_b$ per particle of the YAB model is
\begin{equation}\label{Fb}
\beta \frac{F_b}{N} = 2 \ln X_B + \ln X_A - X_B - \frac{X_A}{2} + \frac{3}{2}
\end{equation}
where $X_A$ and $X_B$ are the fractions of non-bonded patch of each species respectively\cite{Jackson1988} and $\beta = \frac{1}{k_BT}$. For the $YAB$ model they are: 
\begin{equation}
X_A = \frac{1}{1+2\rho \Delta X_B}; \ \ \  X_B = \frac{1}{1+\rho \Delta X_A},
\label{eq:X}
\end{equation}
with $\Delta = v_B [e^{\beta \epsilon_0} - 1] \frac{1 - A\eta - B\eta ^2}{(1 - \eta)^3}$,  $v_B = \pi \delta^4 \frac{15 \sigma + 4 \delta}{30 \sigma^2}$, $A = \frac{5}{2} \frac{3 + 8 \delta/\sigma + 3(\delta/\sigma)^2}{15 + 4\delta/\sigma}$, $B = \frac{3}{2} \frac{12 \delta/\sigma + 5 (\delta/\sigma)^2}{15 + 4\delta/\sigma}$, $\eta =  \frac{\pi}{6} \rho \sigma^3$\cite{Bianchi2006,Bianchi2008}.

The reference HS system must be chosen according to the nature of the molecule.  For non-spherical molecules, the HS reference system effective diameter is not known and needs to take into account correctly the excluded volume of the particles. This is established from the comparison to experiments. Once this is known, experimentally accessible quantities such as the osmotic compressibility of the system can be directly calculated from the expression of $F$. From Eq.~\ref{eq:X} we can calculate the expressions for the fractions of non bonded patches in terms of density and temperature, as
\begin{eqnarray}
X_A &=& \frac{2}{1+\rho\Delta+\sqrt{\rho^2\Delta^2+6\rho\Delta+1}};\nonumber\\  
X_B &=&\frac{\rho\Delta-1+\sqrt{\rho^2\Delta^2+6\rho\Delta+1}}{4\rho\Delta}.
\label{eq:Xtrue}
\end{eqnarray}
Instead of using these two variables, it is more convenient to refer to the so-called bond probability $p$, defined as 
\begin{equation}\label{p}
p \equiv p_B = 1 - X_B = \frac{p_A}{2} = \frac{1 - X_A}{2}.
\end{equation}

While Wertheim theory directly provides overall quantities such as the compressibility or the fraction of bonded $A$ and $B$ groups as a function of $p$, it does not yield the resulting cluster size distribution that would be needed for a comparison with other experimental quantities. We thus apply hyperbranched polymer theory (HPT) exploiting the fact that the YAB molecule is of the kind $AB_{f-1}$ in HPT language with $f=3$\cite{Rubinstein2003}. We consider $p$ to be the fraction of bonded $B$ groups (i.e. the bond probability of Wertheim theory defined above) and $(f-1)p$ the fraction of bonded $A$ groups. There is one non-bonded $A$ group for each cluster, therefore the average number of monomers per cluster is the reciprocal of the fraction of unreacted $A$ groups. The only input then needed to evaluate the cluster size distribution $n(s)$ is the bond probability $p$, which we get from Wertheim theory. In the YAB model, calling $p$ ($2p$) the fraction of $B$ ($A$) patchy sites, the cluster size distribution in the framework of hyperbranched polymer theory is finally given by
\begin{equation}\label{ns}
n (s) = \frac{(2s)!}{s!(s+1)!} p^{s-1}(1-p)^{s+1},
\end{equation}
where $n(s)$ is the probability of finding clusters of size $s$ for a system with bond probability $p$. From the cluster size distributions calculated we then calculate the weight average, the $z$-average and the polydispersity of the clusters for each concentration.

Finally in order to evaluate $S(0)$ as a function of concentration, we simply perform a double derivative of the free energy in order to get the compressibility of the system\cite{Tavares2009}.

\subsection*{Model fitting}
All data are fitted using the orthogonal distance regression procedure\cite{boggs1990,boggs1992}, which includes the experimental errors in both the x and y directions. The full set of fit parameters are given in table \ref{tab:fitresults}. For the hard sphere model fits shown in Fig.~\ref{fig:final} two parameters are fitted: The bonding energy and a scaling factor of the SLS data. The scaling factor is introduced to correct for any errors in the calibration linking the scattering intensity to the apparent molecular mass, and should be close to one. For the sticky hard sphere model three parameters are fitted: The bonding energy, the stickiness parameter and a scaling factor for the SLS data. For the fits using the compressibility from Wertheim theory (figure \ref{fig:S0}), the first fit uses three free parameters, the bonding energy, a scaling factor of the SLS data and an effective hard sphere scaling factor. In the second fit the effective hard sphere scaling factor is locked to $\sigma_{HS}$ and only the bonding energy and scaling factor are fitted. 

\begin{table}[tbh]
			\begin{tabular}{c|rrrr}
	\!\!\!	Model fit \!\!\!\!\!\!& $\epsilon/k_BT$ \!\!\!\!\!\!& $\sigma_{HS}/\sigma$ \!\!\!\!\!\!& $\tau$ \!\!\!\!\!\!& SLS scale \\
		\hline
		\begin{tabular}{@{}c@{}}Compressibility \\ full fit\end{tabular}\!\!\! \!\!\!& $12.04\pm 0.04$  \!\!\!\!\!\!& $2.70 \pm 0.03$  \!\!\!\!\!\!& - \!\!\!\!\!\!& $0.994 \pm 0.002$  \\

		Compressibility \!\!\!\!\!\!&  $12.27 \pm 0.03$\!\!\!\! \!& $2.9^*$  \!\!\!\!\!\!& -\!\!\! \!\!\!& $0.984 \pm 0.003$\\
		HS Model\!\!\!\! \!& $12.33 \pm 0.03$\!\!\! \!\!\!& $2.9^*$\!\!\! \!\!\!&- \!\!\!\!\!\!&  $0.992  \pm 0.008$ \\
		Sticky HS Model  \!\!\!\!\!\!& $12.24 \pm 0.03$\! \!\!\!\!& $ 2.9^*$\!\!\!\! \!& $ 2.45 \pm 0.45$ \!\!\!\!\!\!&  $0.993 \pm 0.006$
		
	\end{tabular}
	\caption{
	\textbf{Table of the obtained fit parameters}. The $^*$ indicates a fixed parameter.}
	\label{tab:fitresults}
\end{table}

\subsection*{Generation of Antibody Clusters}

The antibody clusters were generated using a molecular model of the antibody, from low concentration SAXS data [Article in preparation] constructed using the SAXS modeling software BUNCH\cite{Petoukhov2005}. The FC and FAB domains (generated as described in the \textit{isosurface calculations} section)  were represented as rigid bodies, linked together with a flexible linker of dummy residues. The linker was further constrained by linking the dummy residues that represent the cysteine residues together to simulate the cysteine bridges in the hinge region of an IgG4.

To generate a self-associated antibody cluster containing N antibodies, the following procedure was used:
\begin{enumerate}
	\item An initial antibody was placed with its center of mass (CM) at the origin and oriented randomly.
	\item A new antibody is placed at a distance of 12.5 nm in a random direction from the initial antibody CM, and oriented randomly and a connection between them is recoded.
	\item From the already placed antibodies one is selected at random if it has less that 3 connections.
	\item A new antibody is now placed distance of 12.5 nm in random direction from the selected antibody, oriented randomly. The distance between the CM of the newly placed antibody and all other placed antibodies is calculated, and if it is 12.5 nm or more the connection is recorded. If not step 4 is repeated.
	\item Steps 3 and 4 are repeated until N antibodies have been placed.
\end{enumerate}
For each association number, N, 100 clusters were produced using the method above.

\section*{Acknowledgments}
This work was financed by the Swedish Research Council (VR; Grant No. 2016-03301), the Faculty of Science at Lund University, the Knut and Alice Wallenberg Foundation (project grant KAW 2014.0052), the European Research Council (ERC-339678-COMPASS and ERC-681597-MIMIC) and the European Union (MSCA-ITN COLLDENSE, grant agreement No. 642774).

\bibliography{pnas-sample}

\end{document}